\documentclass[letterpaper, 10 pt, conference]{ieeeconf}  

\IEEEoverridecommandlockouts                              

\usepackage[pdftex]{graphicx}
\usepackage{amsmath}
\usepackage{amsfonts}
\usepackage{graphicx}
\usepackage{mathtools}
\usepackage{caption}
\usepackage{epstopdf}
\usepackage{color}
\usepackage{hyperref}

\title{\LARGE \bf
IoT-Enabled Distributed Data Processing for Precision Agriculture
}

\author{Grigore Stamatescu, Cristian Dr\u agana, Iulia Stamatescu, Loretta Ichim and Dan Popescu
\thanks{*This work has been funded by the Romanian Space Agency (ROSA), through the project "Integrated Multi-Agent Aerial Robotic System for Exploring Terrestrial Regions of Interest" (MAARS), contract no. 185/2017.} 
\thanks{The authors are with the Department of Automatic Control and Industrial Informatics, University Politehnica of Bucharest, 060042 Bucharest, Romania. Grigore Stamatescu is also with the Institute of Technical Informatics, Technical University of Graz, 8010 Graz, Austria
        {\tt\small gstamatescu@tugraz.at}}%
}

\begin{document}

\maketitle
\thispagestyle{empty}
\pagestyle{empty}

\begin{abstract}

Large scale monitoring systems, enabled by the emergence of networked embedded sensing devices, offer the opportunity of fine grained online spatio-temporal collection, communication and analysis of physical parameters. Various applications have been proposed and validated so far for environmental monitoring, security and industrial control systems. One particular application domain has been shown suitable for the requirements of precision agriculture where such systems can improve yields, increase efficiency and reduce input usage. We present a data analysis and processing approach for distributed monitoring of crops and soil where hierarchical aggregation and modelling primitives contribute to the robustness of the network by alleviating communication bottlenecks and reducing the energy required for redundant data transmissions. The focus is on leveraging the fog computing paradigm to exploit local node computing resources and generate events towards upper decision systems. Key metrics are reported which highlight the improvements achieved. A case study is carried out on real field data for crop and soil monitoring with outlook on operational and implementation constraints.

\end{abstract}

\section{INTRODUCTION}

Internet of Things (IoT) systems are based on distributed sensing, computing and communication devices that collaborate in order to monitor and control physical processes. These enable the collection of real world data at an unprecedented scale and resolution which can then be used to improve the models that define the understanding and help the forecasting of the processes, be it technical, social or environmental. New data processing infrastructure are thus needed to store and retrieve the information collected in an online manner while providing mechanisms to run the analysis and control algorithms based on this data. Beyond conventional environmental monitoring as initial key driver of IoT design, current domains include (smart) cities, industry and agriculture. Finally the outcomes of the analysis are either handled in closed loops for control actions or they are supplied to hierarchical entities for decision support.

Among the applications areas mentioned above, precision agriculture represents one of the salient areas where IoT-enabled systems can improve the quality, productivity and increase automation \cite{8355152}. Main challenges in this field relate to reducing input use: water, fertiliser, work, and obtaining better crop yields which is demanded by the market to keep food costs low under the strains of increasing global population. By having access to reliable, on-line information, relayed over distributed networks, domain specialists can oversee tangible improvements \cite{8372905}.

The conceptual and practical challenges that we approach in the design of such systems is related to efficient data reduction and management which impacts directly the congestion and energy metrics of the deployed network. This is performed by proposing a hierarchical data processing architecture in accordance to fog computing design principles. Fog computing as a concept has initially emerged as a computing organisation alternative to leverage intelligent network edge devices which make up modern IoT systems \cite{anawar2018fog}. The limited computing resources available on these embedded devices are thus exploited to reduce the large quantities of collected data and transmit only higher level information pieces upstream. Given the large heterogeneity the processing primitive can run of the edge nodes range from basic threshold detection and averaging up to more advanced outlier detection and embedded learning algorithms.
Wireless sensor networks (WSN) are an enabling technology to deploy fog computing systems \cite{8394851, 8679064} where hundreds to thousands of sensing nodes self organise intro and communicate over low power radio channels. As with the case with agriculture, large areas can thus be covered with multi-hop communication networks as the networking protocols rely on cluster heads, gateways and hubs serving as intermediary data concentrators.
One alternative definition presents fog systems in opposition or as complementary to conventional centralised and large scale cloud infrastructures. The complex functionality of the cloud platform is broken down at the field level over functional or spatially distributed entities which collaborate to achieve a common monitoring, event-detection and control case. In the precision agriculture use case this can help implement an optimised distributed irrigation or fertiliser dosage schemes accounting for local properties and variance of soil, micro-climate and crop particularities. The need to integrate fog computing with cloud computing in this particular scenario lays with the fact that joint observations can be derived when federating high-level information across multiple farms.

The main novelty of the paper is justified by the application of fog computing data aggregation and modelling primitives in the context of IoT-enabled smart agriculture, a highly active area of research currently. The subsequent contributions of the paper can be argued:
\begin{itemize}
\item system architecture for hierarchical data processing and analysis based on field level IoT devices;
\item data aggregation methodology based on the fog computing paradigm under precision agriculture constraints.
\end{itemize}

\section{RELATED WORK}

In \cite{guardo2018fog} a fog computing framework for precision agriculture is introduced. The two tiered system is able to reduce significantly the data transmitted in the network. Reducing the computational loads, and most important, the cloud computing costs associated with centralised processing is highlighted as an essential benefit of the fog approach. The authors of \cite{8521668} propose a hybrid IoT for smart farming in rural areas. The communication network uses 6LoWPAN local radio for the field interfaces while long range connections are implemented over WiFi. A 6LoWPAN border router and dedicated gateway are used to assure cross-domain integration of the networks from field level, intermediate long range relays and cloud. Network requirements for smart agriculture applications are also discussed in terms of throughput, latency and mobility support. These offer a good reference to quantify the data aggregation potential in conjunction with the sensing and control requirements. A distributed computing architecture is presented in \cite{ferrandez2018precision} which the agricultural system basic components such as: crop, soil, climate, water and nutrients, energy. The messaging system is standardised around the Message Queuing Telemetry Transport (MQTT) to interlink sensors, actuators, communication nodes, devices and subsystems \cite{7845442}. A decision tree is designed for irrigation control and integrated on the edge devices for in situ decision making. At the top level cloud services supply data through an end-user dashboard for high level decision support. 

\cite{kamienski2019smart} introduce an intelligent irrigation system based on distributed sensor using the LoRA long range, low rate, nodes and gateways. The FIWARE infrastructure is leveraged as data management middleware platform which provides the support services. Several operation scenarios are discussed based on the scalability requirements, in terms of tens of thousands of nodes. Reference computational resource assessment for cpu, memory and network is also reported. Large scale IoT monitoring is discussed in \cite{popescu2018collaborative}. The focus is on the ground level clustering mechanisms that support the timely collection of data and generating of the field level monitoring events. Aerial robotic platform support is provided through suitable high level control of trajectories for data collection and backhaul. Data reduction is achieved by thresholding over locally computing moving averages in conjunction with expert knowledge adapted to the monitored processes. Several radio access technologies are available to achieve reliable transmissions \cite{7090210}.

\section{SYSTEM ARCHITECTURE AND METHODOLOGY}

\subsection{SYSTEM ARCHITECTURE FOR DATA COLLECTION AND PROCESSING}

\begin{figure*}[h]
	\centering
	\captionsetup{singlelinecheck=false, justification=centering}
	\includegraphics[width=0.7\paperwidth]{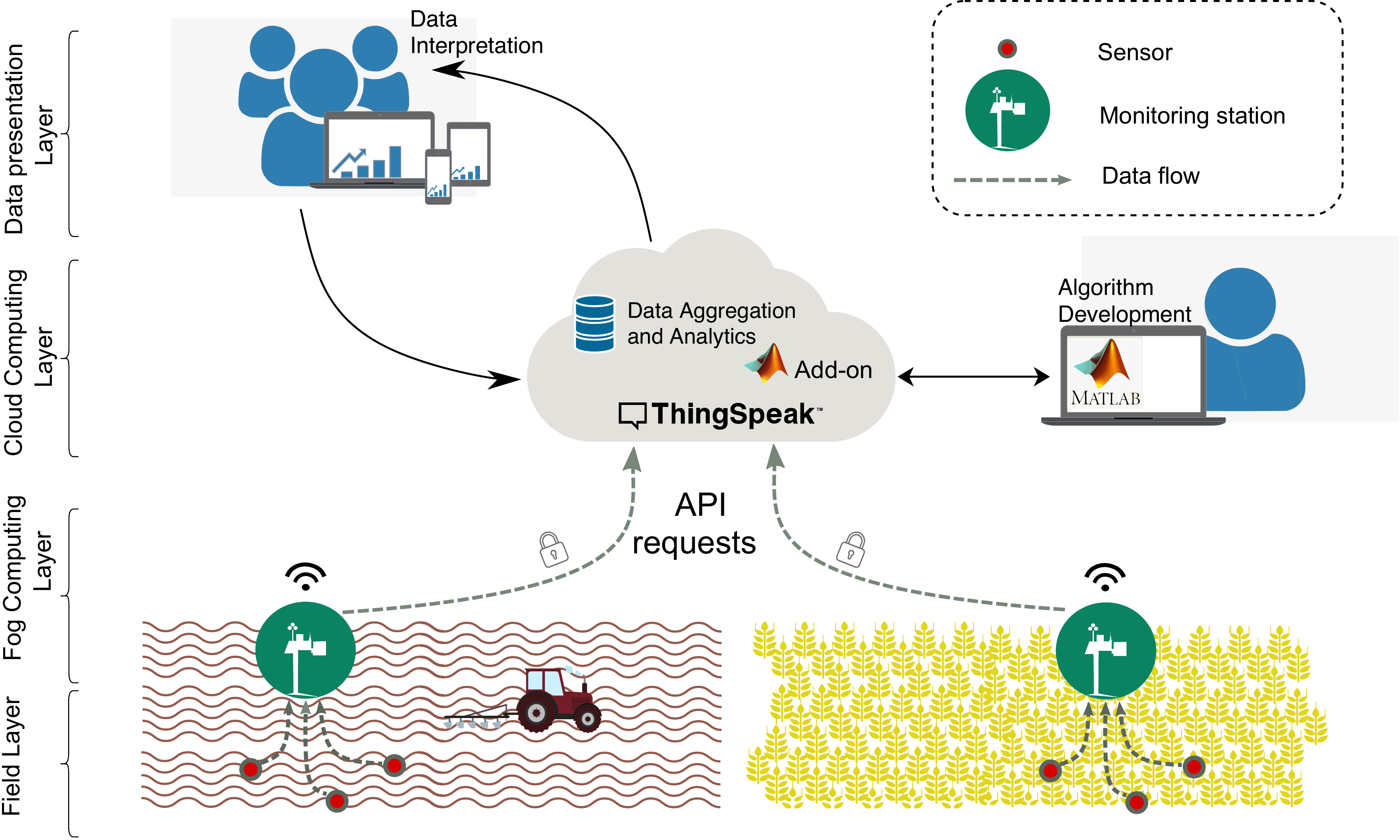}
	\caption{Distributed data processing based on fog computing for precision agriculture}
	\label{fig:perf1}
	\end{figure*}

The proposed system architecture that we have designed for the purpose of efficient data collection and processing in precision agriculture is illustrated in Figure \ref{fig:perf1}. It consists of the following information and physical layers: field layer, fog computing layer, cloud computing layer, data presentation layer, which are linked by cross-layer upstream and downstream data and control information flows. The layer functionality is detailed next:

\begin{itemize}
\item Field layer: includes the actual sensors deployed in the precision agriculture application to measure the physical parameters of interest; these include air temperature, air humidity, solar radiation, soil temperature at various depths, windspeed and rainfall; the field layer can also be expanded to accommodate intelligent actuators e.g. for irrigation or fine grained nutrient dosage, to execute commands incoming from higher level systems;
\item Fog Computing layer: the fog nodes collect data from the sensors and run the data processing primitives for intelligent aggregation in order to reduce network traffic and energy expenditure; the main idea is to locally derive basic model characteristics of the particular process which are sent to the cloud in compact form; correlations between the sensed variables can also be exploited at this level for local decisions thus avoiding completely the increased cost and latency of the upper layers; 
\item Cloud computing layer: data is streamed towards a common cloud platform; regarding the particular implementation we use the ThingSpeak \cite{thingspeak} platform in conjunction with Matlab algorithm development for higher level processing routines; at the cloud layer the model parameters allow the reconstruction of the time series characteristics if needed, while accounting for the inherent modelling errors;
\item Data presentation layer: is concerned with the front-end software systems that present the outcomes of the data analysis to end-users or decision makers with the ability to provide mobile access and timely alerts in the case of event detection; parametrisation of the process by domain experts is also achieved at this layer.
\end{itemize}

\begin{figure}[htb]
	\centering
	\includegraphics[width=\columnwidth]{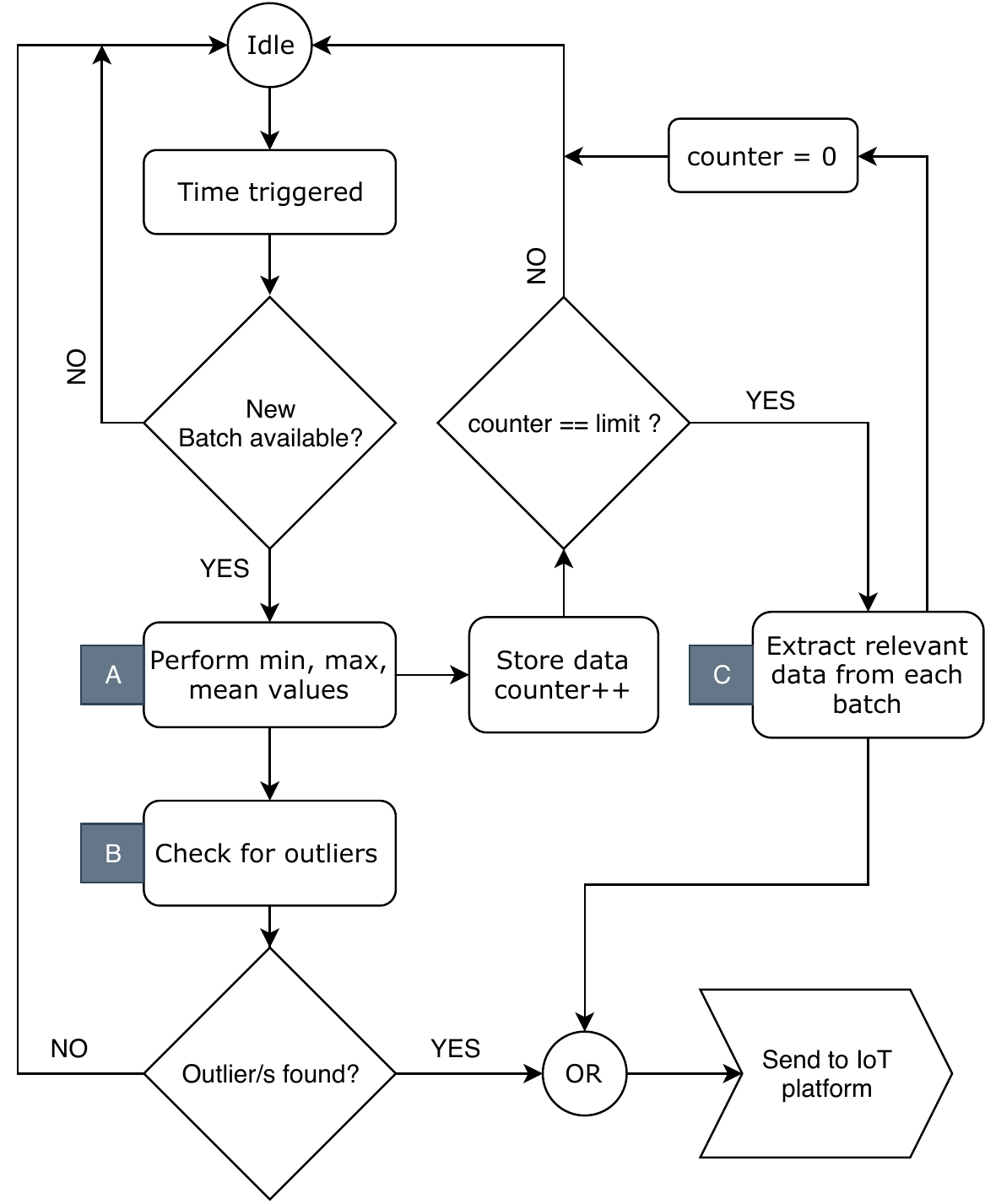}
	\caption{Fog Computing algorithm}
	\label{fig101}
\end{figure}

A more detailed algorithm flowchart is provided in Figure \ref{fig101}. It includes the steps for algorithm description which runs on the fog computing node.

In-field measurements are uploaded to the IoT application in two ways depending on the type of information: events and measurements. Note that, a primary batching procedure is usually available for most of the monitoring systems, basically consisting of performing minimum, maximum and mean value during a specific period of time. We consider this as the starting point for further local data processing.

\emph{Primary batch aggregation} Note that, a primary batching procedure is usually available for most of the monitoring systems, basically consisting of performing minimum, maximum and mean value during a specific period of time. We consider this as the starting point for further local data processing.

For instance, batches are defined within 30 minutes. Once a new batch is available, $min, max$ and $mean$ values are computed (step A).

\emph{Check for outliers procedure} For each batch of measurements, an outliers' check procedure is performed, considering an acceptance bandwidth of data variance for the measured value around the mean (step B). The procedure outputs an event if the minimum or maximum values exceeds the thresholds. The event $E$ is defined as:

\begin{equation}
E=\left \{ e(x_{i})\in Q, T_{min}<x_{i}<T_{max} \right \}
\end{equation}

where:

\begin{itemize}
	\item $x_{i}$ is the measured value at iteration $i$
	\item $ T_{min}$ and $ T_{max}$ are thresholds computed as:
	\begin{equation}
		T_{min}=mean(1-w)	
	\end{equation}
	\begin{equation}
		T_{max}=mean(1+w)	
	\end{equation}
	where
	$w$ is a weight for acceptance bandwidth size define.
\end{itemize}

\emph{Relevant data extraction} Aggregated data sets are achieved based on different methods. All seek for relevant data point, aiming to a reduced size set providing at the same time a satisfying reconstruction of the initial data.

One effective method, in terms of data volume, is based on using the $min$ and $max$ values extraction, computed for 24 hours. 
It is obvious that this method is suitable only for measurements that follow a regular shape during time, with insignificant variations during a day. A measurement for which this method is suitable is the soil temperature. 

Instead, change detection is a common method applicable for irregular shaped data sets. This method follows extraction of data points where trend changes occur.

Given a set of data point $(x_{i}, y_{i}), i = 1, ..., n$, trend $t_{i}$ is followed for each pair  $x_{i}, x_{i+1}$, such that for

\begin{equation}
\begin{aligned}
 x_{i+1}- x_{i}>\delta \implies t(i)=1\\
 x_{i+1}- x_{i}<\delta \implies t(i)=-1\\
  x_{i+1} = x_{i} \implies t(i)=0
  \end{aligned}
\end{equation}

Then, if $t(i)\neq t(i+1)$ means that a trend change is detected. The coresponding data point $x_(i+1)$ is  added to the relevant data set.

Relevant data extraction (step C) is performed when a set of primary aggregated batches is available. 

\subsection{DATA AGGREGATION}

One reference method of extracting high level information from sensor data is Symbolic Aggregate Approximation (SAX) \cite{Keogh:2005:HSE:1106326.1106352}. It operates by assigning label symbols to segments of the time series thus porting it in a unified lower dimension representation. It belongs to the family of time series data mining techniques leading to non-parametric modelling. Ranges are identified through the data histogram or in a uniform manner. The method provides linear complexity and opens up the use and application of multiple statistical learning tools. Parametrisation of SAX is highly important by defining the number of segments and the alphabet size which can influence the quality and robustness of the result.

The background on which SAX has been defined is established by PAA \cite{Chakrabarti:2002:LAD:568518.568520} where symbols are attributed to the aggregated numerical values listed by PAA. Several discrete event models can incorporate the resulting aggregated segments e.g. Markov models in order to compute the probability of the observed patterns for future observations. According to the PAA method description, starting with a time series $X$ of length $n$, this is approximated into a vector $\bar{X}=(\bar{x}_1,...,\bar{x}_M)$ of any length $M\leq n$, with $n$ divisible by $M$. Each element of the vector $\bar{x}_i$ is calculated by: 

\begin{equation}
\bar{x}_i=\frac{M}{n}\sum^{(n/M)i}_{j=n/M(i-1)+1}x_j
\end{equation}

The dimensionality of the time series is thus reduced from $n$ to $M$ samples by initially dividing the original data into $M$ equally sized frame and then compute the mean values for each frame. A new sequence is achieved by putting the mean values together which is considered to be the PAA transform (approximation) of the original data. With regard to computational considerations, the PAA transform complexity can be reduced from $O(NM)$ to $O(Mm)$ with $m$ being the number of frames as tuning parameter of the method. The distance measure between two time series vector approximations $\bar X$ and $\bar Y$ is defined as:

\begin{equation}
D_{PAA}(\bar X,\bar Y)=\sqrt{\frac{n}{M}}\sqrt{\sum^M_{i=1}(\bar x_i - \bar y_i)}
\end{equation}

It has been shown by the proposers of the method that PAA satisfies the lower bounding condition and guarantees no false dismissals such that:

\begin{equation}
D_{PAA}(\bar{X},\bar{Y}) \leq D(X,Y)
\end{equation}

\subsection{INTERPOLANT METHODS}

The Cloud-based application rebuilds data sets by estimates based on interpolation mechanisms. For performance evaluation we showcase three methods: the common linear interpolant (also referred as piecewise linear interpolant ) and two closely related interpolants, cubic  \textit{spline} and shape preserving \textit{Piecewise Cubic Hermite Interpolating Polynomial (pchip)}.

Given a set of data points $\left ( x_{i}, y_{i} \right ), \left ( x_{i+1}, y_{i+1} \right ), ..., $ $\left ( x_{n}, y_{n} \right )$, the linear interpolation is defined as the concatenation of linear interpolants between each pair of data points, thus a set of straight lines between each data points. Any pair of data points with   $x_{i}\neq x_{i+1}$ determines a unique polynomial $p$ of degree less than two whose graph passes through the two points with the property:

\begin{equation}
	p(x_{i})=y_{i}
\end{equation}

with the form:

\begin{equation}
	p(x)=a_{1}x+a_{0}
\end{equation}

a  1-D linear interpolation.

In general, given $n$ points $\left ( x_{i}, y_{i} \right ), i = 1, ..., n$, with disting  $x_{i} $, a polynomial of degree less than $n$ whose graph passes through the $n$ points denoted $P_{n}(x)$, is expressed in the \textit{Lagrange} form as:

\begin{equation}
          P_{n}(x)=\sum_{i=1}^{n}\Bigg(\prod_{\begin{matrix}
{\scriptstyle j=1}\\
{\scriptstyle j\neq i}\\ 
\end{matrix}}^{n}\frac{x-x_{j}}{x_{i}-x_{j}}\Bigg)y_{i}
\label{poly_lag}
\end{equation}

The Lagrange form in ~\eqref{poly_lag} can be written out in power form of an interpolating polynomial as,

\begin{equation}
          P_{n}(x)=a_{1}x^{n-1}+a_{2}x^{n-2}+ ... +a_{n-1}x+a_{n}
\label{poly_pow}
\end{equation}

where the coefficients $a_{k}$ are computed through a system of linear equations:
\begin{equation}
\begin{bmatrix}
x^{n-1}_{1} & x^{n-2}_{1} & ... & x_1  & 1\\ 
x^{n-1}_{2} & x^{n-2}_{2} & ... & x_2  & 1\\
\vdots & \vdots  & \vdots  & \vdots   & \vdots \\  
x^{n-1}_{n} & x^{n-2}_{n} & ... & x_n  & 1\\
\end{bmatrix}
\begin{bmatrix}
a_{1}\\
a_{2}\\
\vdots\\
a_{n}
\end{bmatrix}=
\begin{bmatrix}
y_{1}\\
y_{2}\\
\vdots\\
y_{n}
\end{bmatrix}
\end{equation}

Considering this, a piecewise linear interpolant is produced by first computing the divided difference:

\begin{equation}
         \delta_{i}:=\frac{y_{i+1}-y{i}}{x_{i+1}-x{i}}
\label{div_dif}
\end{equation}

Then the interpolant is constructed as:

\begin{equation}
         P(x)=y_{i}+\delta_{i}(x-x_{i})
\end{equation}

Further, for piecewise cubic polynomials, considering an interval $x_{i}\leq x\leq x_{i+1}$ let $h_{i} := x_{i+1}-x{i}$ be the length of an $i^{th}$ interval and $d_{k}:=P'(x_{i})$.
Therefore, using this derivative it is possible to adjust the interpolant in order to enforce smoothness, by forcing the pair of derivatives from consecutive piecewice cubics to agree.

All piecewise cubic hermite interpolating polynomials are continuous and have a continuous first derivative.
In particular, \textit{spline} is oddly smooth, meaning that it's second derative also varies continously. 

Instead, \textit{pchip} is not as smooth as  \textit{spline}, it is actually designed so that it never overshoots the data. The slopes are chosen so that $P(x)$ preserves the shape of data and also respects monotonicity.

\section{EXPERIMENTAL RESULTS}

We collect experimental data from a network of field devices installed on site at an experimental research farm. Form the long term monitoring dataset we select a sample for analysis that covers one month of data. The data is preprocessed for missing values, noise removal and averaged over 30 minute intervals.

\begin{figure}[htb]
	\centering
	\includegraphics[width=\columnwidth]{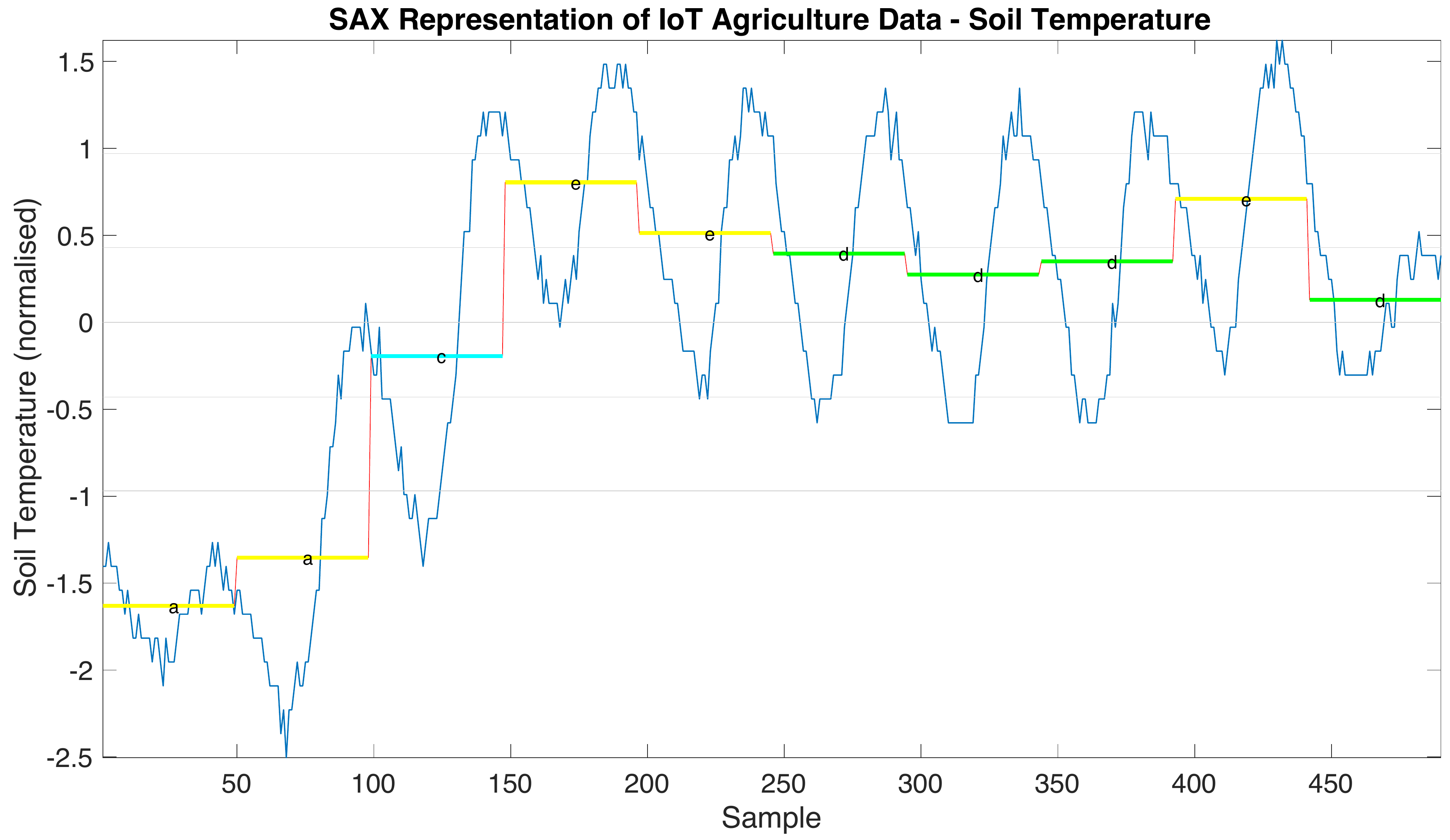}
	\caption{Symbolic Aggregation Approximation - Soil Temperature}
	\label{fig11}
\end{figure}

We first illustrate the application of the SAX method on the measured values for soil temperature and solar radiation in Figure \ref{fig11} and Figure \ref{fig13}. Segment levels codify the evolution of the respective time series and provide a compact representation with considerable impact on the data storage and transmission requirements at the fog node. Finer grained patterns can be observed by zooming in at the daily level as is illustrated in Figure \ref{fig13}. Based on the selected segment labels, if the expected value deviates significantly by entering a different label segment, an event detection primitive can trigger a communication message from the node upstream.

\begin{figure}[htb]
	\centering
	\includegraphics[width=\columnwidth]{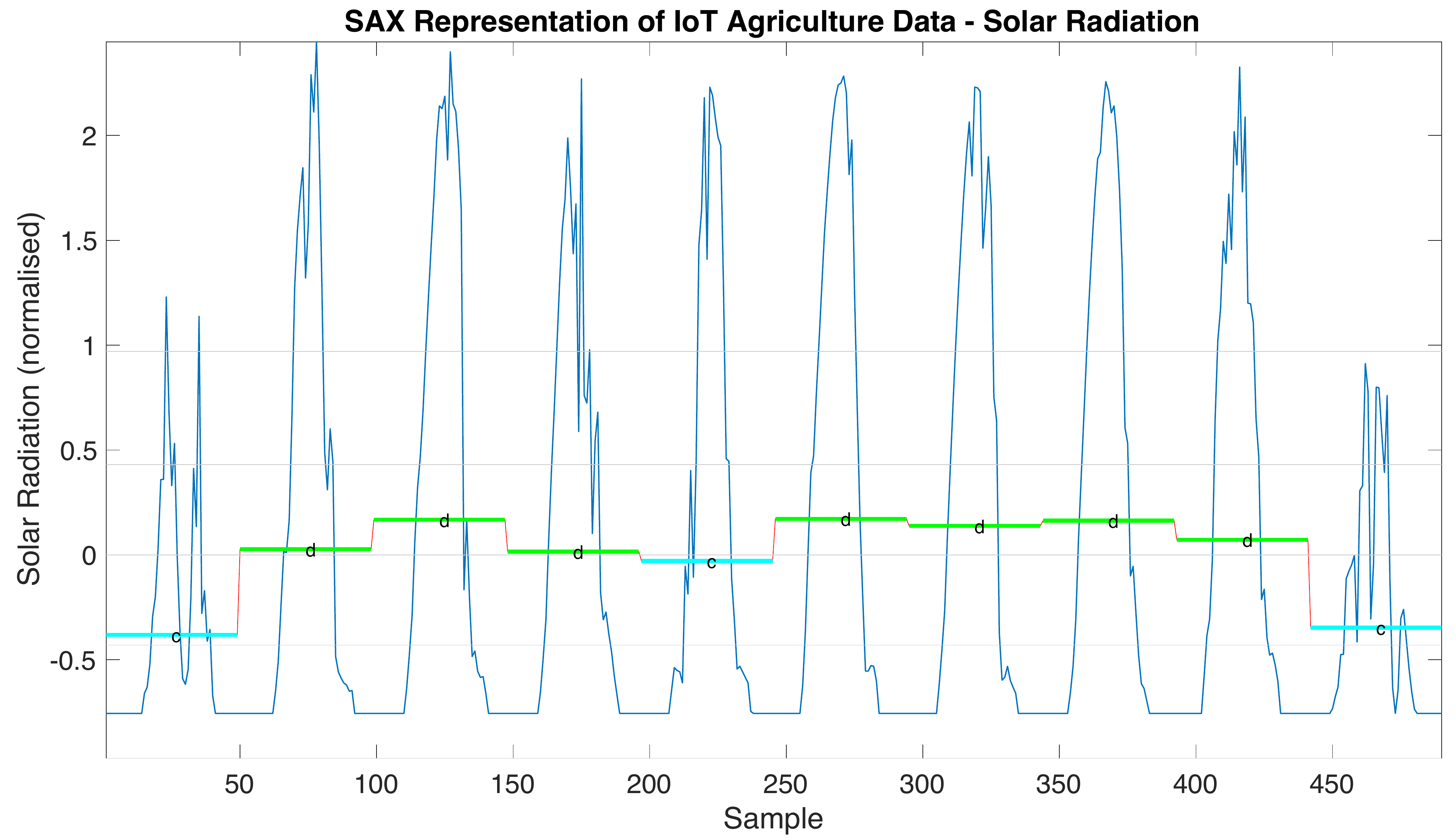}
	\caption{Symbolic Aggregation Approximation - Solar Radiation}
	\label{fig12}
\end{figure}

\begin{figure}[htb]
	\centering
	\includegraphics[width=\columnwidth]{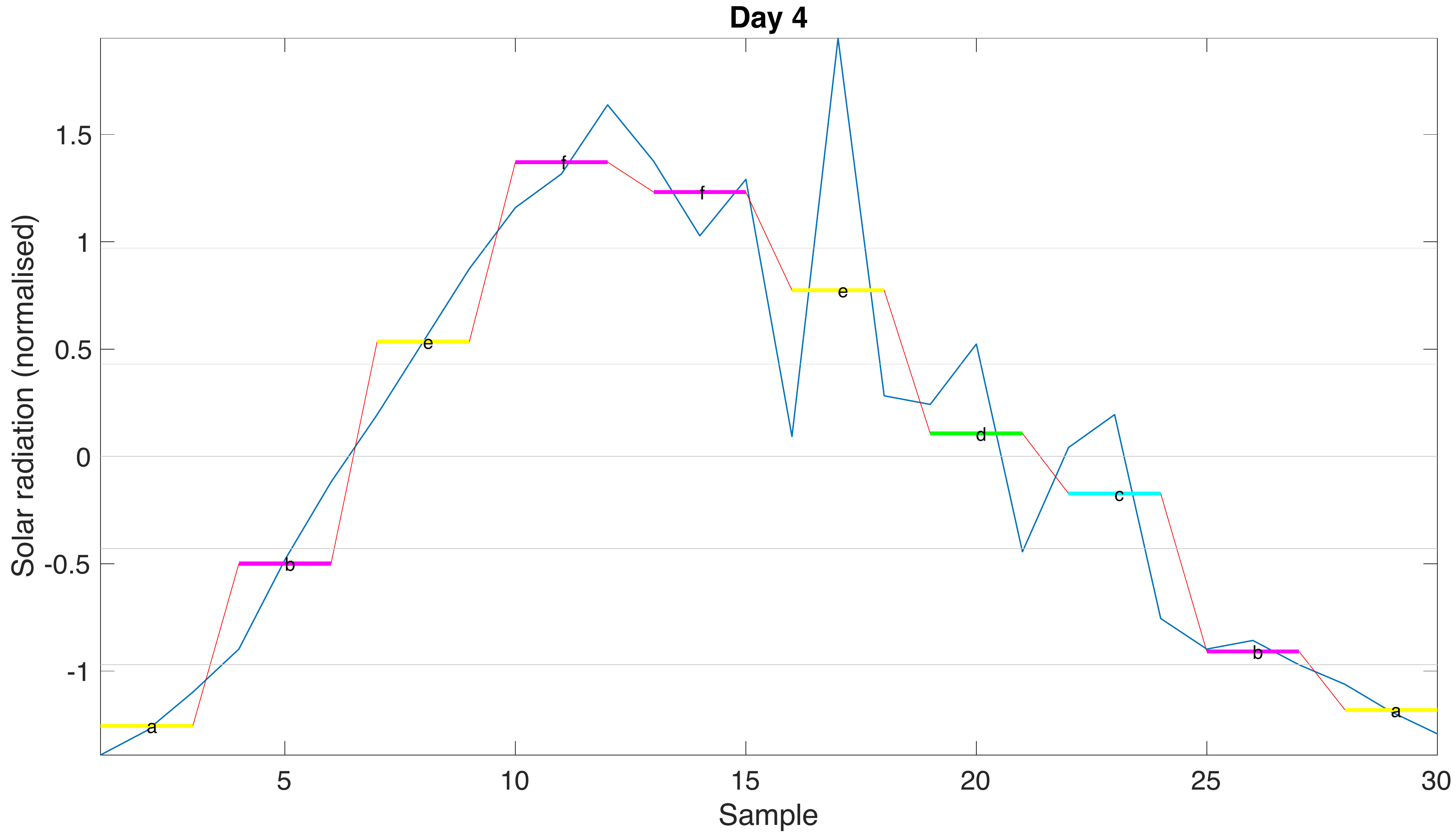}
	\caption{Solar Radiation - Day level aggregation}
	\label{fig13}
\end{figure}

In order to evaluate reconstructed data consistency, achieved through different estimating algorithms, more precisely the proposed interpolants, some well known goodness-on-fit statistics are performed:
\begin{itemize}
	\item Sum of squares of errors (SSE) - measures the total deviation of the response values from the fit to the response values and is defined as:
	\begin{equation}
		SSE=\sum_{i=1}^{n}w_{i}(x_{i}-P(x_{i}))^{2}
	\end{equation}
	
	where $w_{i}$ is the weight for the $i^{th}$ error between estimated  $i^{th}$ value and the empiric data
	\item R-square - measures how successful the fit is in explaining the variation of the data and is expressed as:
	\begin{equation}
		R-square=1-\frac{SSE}{SST}
	\end{equation}	
		
	where 
	
	\begin{equation}
		SST=\sum_{i=1}^{n}w_{i}(x_{i}-\bar{x_{i}})^{2}
	\end{equation}
	
	where $\bar{x_{i}}$ is the mean value of $x_{i}$ dataset.
	\item Root mean square error (RMSE) - is an estimate of the standard deviation of the random component in the data and is expressed as:
	
          \begin{equation}
		RMSE=\sqrt{\frac{SSE}{n}}
	\end{equation}
	
\end{itemize}

Results are summarised in Table I.

\begin{table}[htb]
 \caption{Goodness-on-fit statistics results}
\label{my-label}
\begin{tabular}{|*{4}{p{1.8cm}|}}
\hline
              & pchip RMSE & sline RMSE & interp1q RMSE  \\ \hline
soil temperature   & 0.0852      & 0.1104         & 0.0795   \\ \hline
solar radiation & 0.2627       & 0.3691      & 0.2551    \\ \hline

\end{tabular}
\end{table}

\begin{figure}[htb]
	\centering
	\captionsetup{singlelinecheck=false, justification=centering}
	\includegraphics[width=\columnwidth]{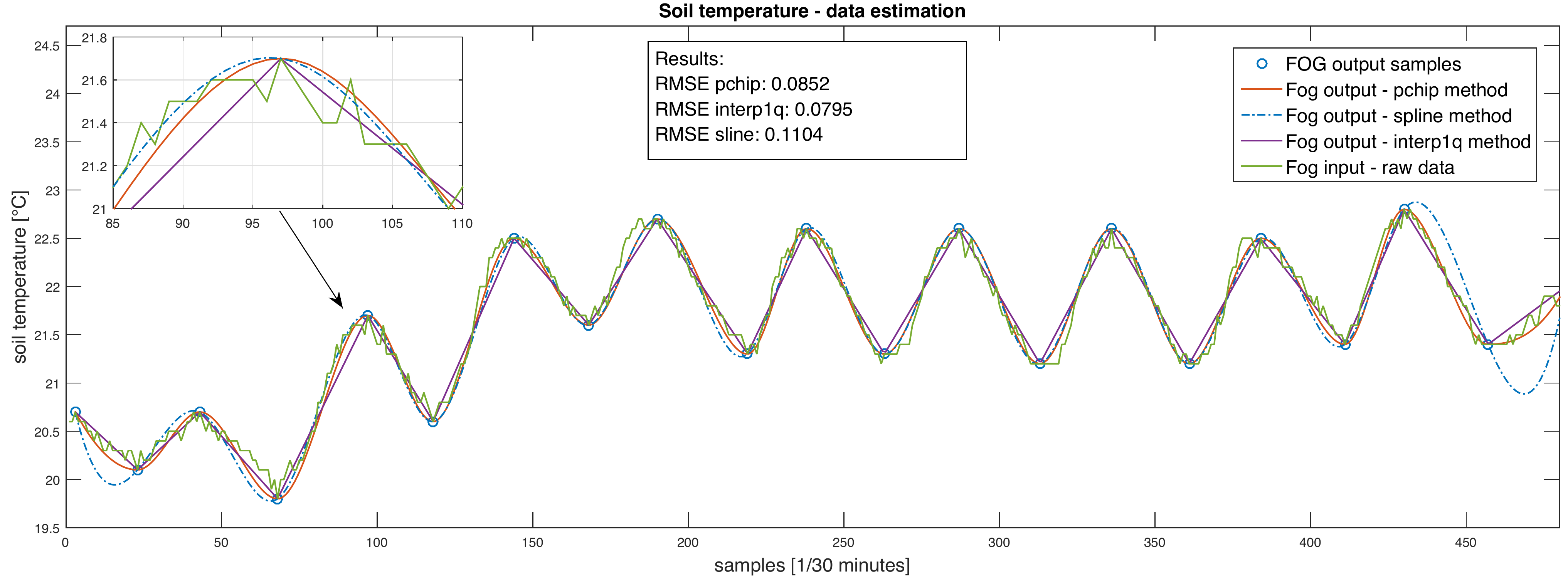}
	\caption{Soil Temperature}
	\label{fig01}
\end{figure}

\begin{figure}[htb]
	\centering
	\captionsetup{singlelinecheck=false, justification=centering}
	\includegraphics[width=\columnwidth]{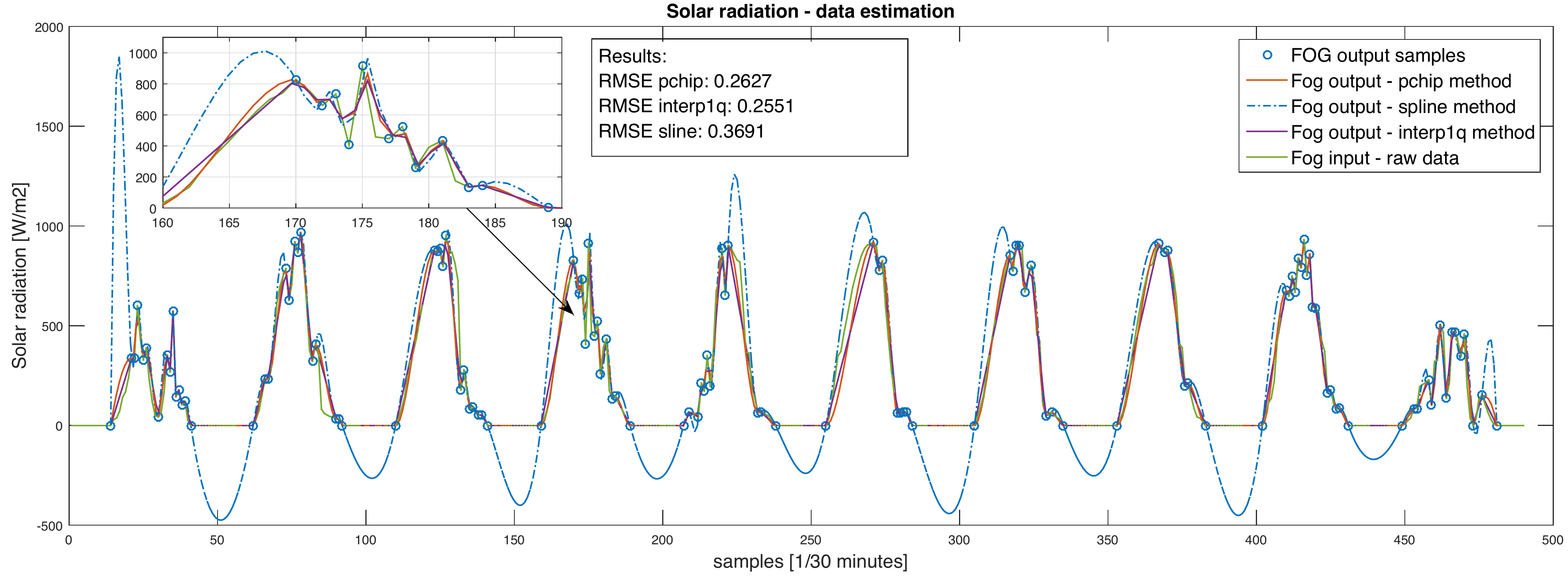}
	\caption{Solar Radiation}
	\label{fig02}
\end{figure}

\begin{figure}[htb]
	\centering
	\includegraphics[width=0.63\columnwidth]{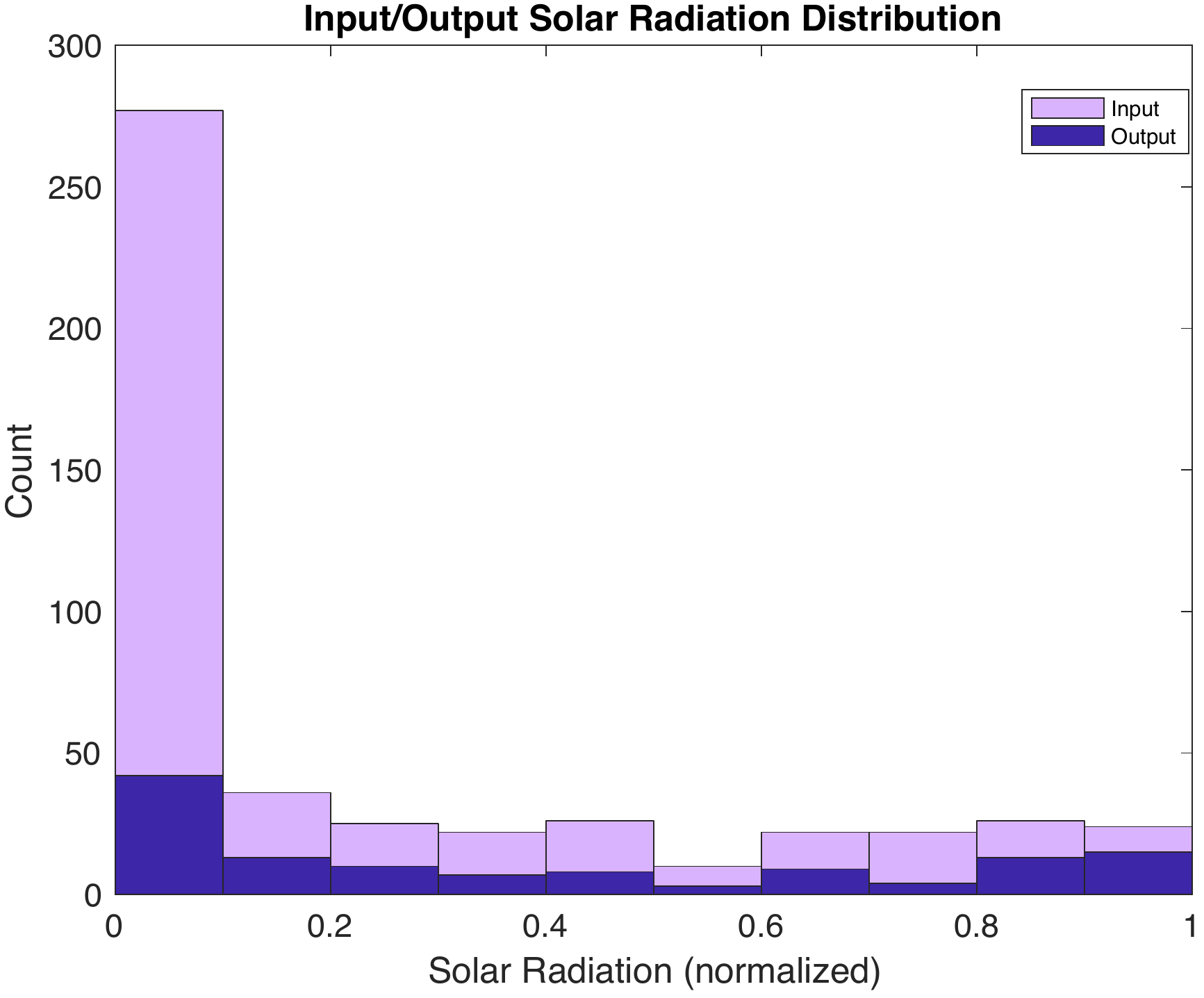}
	\caption{Solar Radiation - Histogram}
	\label{fig013}
\end{figure}

\begin{figure}[htb]
	\centering
	\includegraphics[width=0.63\columnwidth]{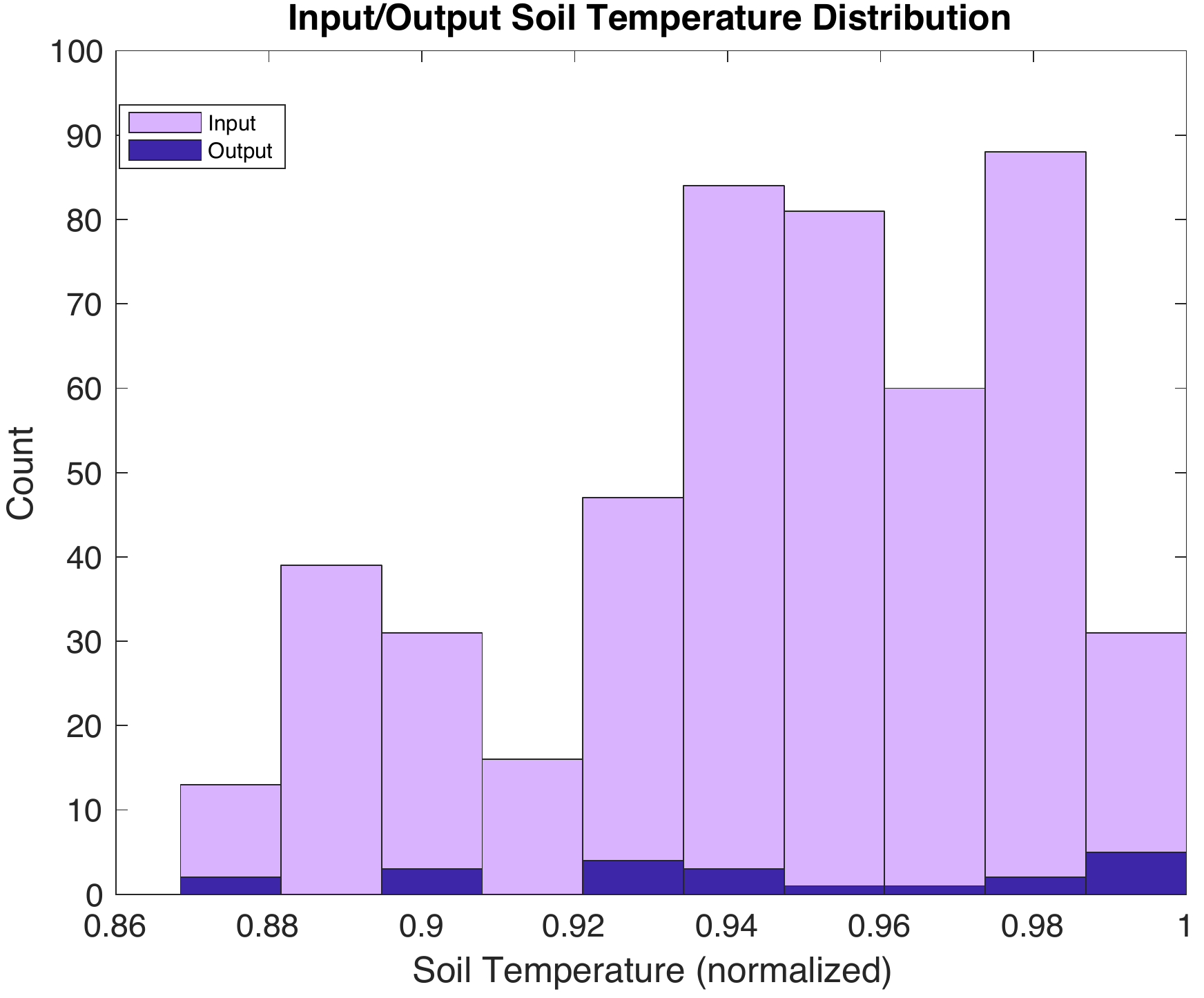}
	\caption{Soil Temperature - Histogram}
	\label{fig014}
\end{figure}

Figure \ref{fig01} and Figure \ref{fig02} graphically depict the results of applying the alternative methods of interpolation on the two time series. In Figures \ref{fig013} and \ref{fig014} the histograms quantify the associated data reduction between the raw input data and the interpolant methods presented.

For this case, the monotonicity property of \textit{pchip} is more desirable than the smoothness property of \textit{spline}, which in some places overshoots the data, thus one may  prefer the good behavior of the shape preserving  \textit{pchip} method. Note that, as with the linear interpolation, when there are two consecutive points with the same value, the interpolant is constant over that interval. This behaviour was expected and it is appropriate in this context.

Even if the metrics indicate better fitting for linear interpolation through the studied cases, one can choose the \textit{pchip} method, given that the results are quite close and it does a much more visual pleasing representation, in particular better modelling the peeks and following the expected behaviour around the baseline. 

\section{CONCLUSIONS}

The paper presented a system architecture and distributed data processing application based on IoT in precision agriculture. By exploiting the dense spatial and temporal distributions of the sensing nodes, intelligent data reduction through aggregation and model reconstruction is illustrated for significants benefits for network congestion and energy efficiency. As the results achieved show promise, future work is focused on extensive evaluation for online decision making by domain experts in order to improve the reconstructed data quality.

\addtolength{\textheight}{-12cm}   



\bibliographystyle{IEEEtran}
\bibliography{IEEEabrv,med19_refs}

\end{document}